# Scaling Expected Force: Efficient Identification of Key Nodes in Network-based Epidemic Models

PAOLO SYLOS LABINI, Free university of Bolzano, Italy

ANDREJ JURCO, Free university of Bolzano, Italy

MATTEO CECCARELLO, University of Padova, Italy

STEFANO GUARINO, Institute for Applied Mathematics "Mauro Picone", CNR, Italy

ENRICO MASTROSTEFANO, Institute for Applied Mathematics "Mauro Picone", CNR, Italy

FLAVIO VELLA, University of Trento, Italy

Centrality measures are fundamental tools of network analysis as they highlight the key actors within the network. This study focuses on a newly proposed centrality measure, Expected Force (EF), and its use in identifying spreaders in network-based epidemic models.

We found that EF effectively predicts the spreading power of nodes and identifies key nodes and immunization targets. However, its high computational cost presents a challenge for its use in large networks. To overcome this limitation, we propose two parallel scalable algorithms for computing EF scores: the first algorithm is based on the original formulation, while the second one focuses on a cluster-centric approach to improve efficiency and scalability. Our implementations significantly reduce computation time, allowing for the detection of key nodes at large scales.

Performance analysis on synthetic and real-world networks demonstrates that the GPU implementation of our algorithm can efficiently scale to networks with up to 44 million edges by exploiting modern parallel architectures, achieving speed-ups of up to 300×, and 50× on average, compared to the simple parallel solution.

Additional Key Words and Phrases: Big-Data, SIR, Epidemic, Expected Force, Graph Centrality, Network, Parallel Computing

## 1 INTRODUCTION

A fundamental problem in the study of epidemic outbreaks is the identification of individuals playing a critical role in the epidemic process. These include so-called "super-spreaders", i.e., individuals that are directly responsible for a great number of infection events, as well as individuals that might bring the epidemic from one sub-population to another, that are likely to infect one or more super-spreaders, or that have frequent interactions with particularly vulnerable people.

In network-based epidemic models, where individuals are represented as vertices in a graph, key actors can be identified by observing their role in simulated epidemic outbreaks: *how many individuals are directly infected by a node? How often does that node seed a global outbreak? Do epidemics grow slower when that node is immunized?* Answering these questions for large-scale networks, however, requires a non-trivial amount of computation, and strong assumptions on the specifics of the spreading process. For this reason, a large amount of research seeks to predict the dynamic of epidemic diffusion from the topological structure of the network [1–5]. In this context, the importance of an individual is gauged by computing suitable graph metrics - "centrality" metrics - over their contact network. Centrality metrics are as varied as the dynamic properties they try to capture. A vertex with a high *degree*, for example, is likely to be a super-spreader, directly infecting many other nodes. On the other hand, a vertex with high *betweenness centrality* - one

Authors' addresses: Paolo Sylos Labini, Free university of Bolzano, Bolzano, Italy, paolo.syloslabini@stud-inf.unibz.it; Andrej Jurco, Free university of Bolzano, , Bolzano, Italy, andrej.jurco@me.com; Matteo Ceccarello, University of Padova, Via Gradenigo 6, Padova, Italy, matteo.ceccarello@unipd.it; Stefano Guarino, Institute for Applied Mathematics "Mauro Picone", CNR, Via dei Taurini 19, Rome, Italy, s.guarino@iac.cnr.it; Enrico Mastrostefano, Institute for Applied Mathematics "Mauro Picone", CNR, Via dei Taurini 19, Rome, Italy, e.mastrostefano@iac.cnr.it; Flavio Vella, University of Trento, , Trento, Italy, flavio.vella@unitn.it.

that belongs to many shortest paths between other vertices - may be a good target for immunization when trying to isolate different regions of the graph [6].

An important property of a node is its *spreading power* during an epidemic. The epidemic spreading power of a node is defined as the size of the infection tree rooted in that node - i.e. the number of infections that are *caused* by that node, either directly or indirectly. This value depends, of course, on the specific dynamics of an epidemic. Yet its average value in small-world network outbreaks, where long infection paths are rare, mainly depends on the local network around the node. This is the underlying idea of the *Expected Force of Infection (EF)* [7], a recently proposed centrality metric which ranks nodes based on the number of different ways in which, once infected, they may lead to other, close nodes catching the disease. Providing a tight link between a dynamic property - the spreading power - and local topology, the EF is a good candidate metric for the study of spreading processes on static networks. However, until now, its high computational costs prevented a systematic evaluation of its potential.

In this work, we explore the use of the Expected Force in the study of *Susceptible-Infected-Recovered* (SIR) spreading processes and provide effective means for its fast computation. Our contribution is articulated as follows:

- we provide a new, efficient algorithm for the Expected Force, along with two scalable, parallel implementations, which allow us to calculate the EF of previously untractable graphs.
- we present a comprehensive analysis of the performance and scalability of parallel EF algorithms, discussing challenges and design choices. Our optimized parallel implementation is up to 300× faster than a parallelized version of the original EF algorithm, allowing us to process a 44 million-edges graph in under 4 minutes. We make our implementation publicly available.
- we conduct a systematic analysis of individual spreading power in SIR epidemics on real-world and synthetic networks. We compare these results with popular centrality metrics, showing that the EF is the best predictor of empirical spreading power.
- we show that seeding nodes with high EF increases the likelihood and size of global outbreaks. Similarly, we show that immunizing nodes with high EF reduces the likelihood and size of outbreaks.

This paper is organized as follows. Section 2 introduces background concepts and relevant works. In Section 3, we present the epidemic model used in our simulations and the centrality measures we investigated. The main contribution of the paper is presented in Section 4, where we discuss parallel algorithms for computing the EF and the implementations. Experiments and the performance analysis are reported in Section 5. Finally, conclusions and future directions highlight the contribution of this work in Section 6.

## 2  BACKGROUND AND RELATED WORKS

The classical compartmental models of disease propagation [8, 9] describe the system at the population-level: only the size of the sub-populations in different health states matter and the transitions between these compartments are governed by constant rates in ordinary differential equations. This approach is often referred to as a mean-field approximation, and it is equivalent to the assumption of *homogeneous mixing, i.e.*, that contacts between any two individuals occur randomly with equal probability, so that a contagious individual may infect any other member of the population with constant probability. While easier to treat analytically, this approach has rapidly proved inappropriate to correctly estimate even macroscopic features of the epidemic and to design counteractions [10]. Network-based models for human interactions were thus introduced, allowing to mimic repeated contacts with the same individuals [11] or to observe the typical hierarchical spread that occurs in real-world scale-free networks [12]. At the node level, it is quite natural to use topological centrality metrics to try and predict which of the vertices will be especially relevant in spreading process [1–5], but the question of which measure can best identify the key spreaders and the best immunization targets remains an open one [5].

In this work, we investigate the use of a recently proposed measure, the *expected force of infection* (EF) [7], specifically designed to capture the spreading power of nodes. The EF measures, in essence, the out-degree of clusters centred on a certain node. A naive implementation of the EF, as originally proposed by Lawyer [7], is computationally intensive. Thus, there is limited research on its ability to identify critical epidemic spreaders.

Lawyer [13] also applied EF to the world airline network (WAN), which plays a central role in the transport network for infectious diseases. In the study, a modified version of the EF successfully identified key players by explaining much of the variation in epidemic outcomes. The approach proposed simulates a pandemic by using the GLEAMviz simulator, which integrates real-world global population and mobility data with an individual based stochastic mathematical model of the infection dynamics [14]. However, the application of the methods was limited by the modest size of the WAN (3458 airports connected by $\sim 70K$ routes).

Calculating most centrality metrics, with the exclusion of the simplest ones such as the degree, is very computationally demanding. For instance, the exact sequential computation of the betweenness centrality requires time $O(nm)$ on a graph with $n$ nodes and $m$ edges [15], which is clearly untenable for large graphs. Therefore, efforts have been made for either running the computation in parallel [16–18] or approximating the result [19, 20].

Our work addresses the exact computation of the Expected Force by making effective use of parallel computing resources. When designing parallel algorithms for graph problems, it is common to either adopt a *vertex-centric* approach [21] or an *edge-centric* approach [22, 23]. *Vertex-centric* approaches often suffer from load unbalancing, due to the skewed degree distributions of real-world networks. However, it is possible to mitigate the load imbalance by concatenating group of edges from the same source indices [23]. Our parallel implementation of the original EF algorithm formulation, which we will use as a baseline, adopts a similar technique. In this work, we expose better opportunities for parallelism by adopting a *cluster-centric* approach that leverages the definition of Expected Force. Thus, instead of parallelizing across vertices or edges, we process the clusters in parallel. Many graph analytics frameworks exploit GPU capabilities by mapping the graph programming model into the computational units in different ways. These choices strongly depend on how we express the computation on the graph (e.g., a single vertex to a thread, or its neighbours to a warp and so forth). For example, Hong et. al proposed a *warp-centric* approach to alleviate the difficulties coming from the irregular nature of the graph and maximize the memory throughput [24]. In this work, we assign multiple clusters into chunks that will be assigned to CUDA streams to exploit their asynchronous behaviour. Other methods related to the data layout and data structures, workload mapping, and GPU programming techniques not strictly related to this work are summarized by Shu [25].

## 3   METHODOLOGY AND MODELS FOR THE EPIDEMIC ANALYSIS

A good centrality metric should be informative of the behaviour of a node during an epidemic. Thus, to demonstrate the efficacy of a centrality measure, we need to compare the centrality value of nodes with quantifiable, epidemic-related properties of a node– e.g., the percentage of global outbreaks seeded by that node. In this section, we discuss how we model and simulate SIR epidemics on a network, which node properties we observe during the simulations, and which centrality metrics we compare with these properties.

### 3.1   Epidemic Model

In this work, the propagation of the disease in the population is modelled through a Susceptible-Infected-Recovered model. At each time step, at a rate equal to $\beta$ an infected individual spreads the disease to susceptible individuals, whereas, at a fixed rate equal to $\gamma$, it recovers from the disease. The classical deterministic SIR model is characterized by threshold behaviour – given an original infected individual (the *index case*), the epidemic process may give rise to an epidemic or become extinct depending on whether the value of the base reproduction number is greater or less than 1. The reproduction number is defined as the expected number of infections directly caused by the index case in a completely susceptible population, $R_0^{\text{index}} = \mathbf{E}\left[\sum_{v \neq u} \frac{\beta}{\mu}\right]$, where $\mathbf{E}$ indicates the average w.r.t. the choice of the index case. With a single infected individual, the fraction of the population involved in each outbreak is expected to follow a bimodal distribution. The first peak of the distribution corresponds to processes that die out in the early stages of the epidemic (*minor* outbreaks), while the second corresponds to genuine outbreaks that involve a large fraction of the population (*global* outbreaks) [26, 27]. As done in [28, 29], we calibrate our SIR model on a common Influenza Like Illness (ILI), setting the average time to recovery to 3 days and $R_0^{\text{index}} = 1.3$, thus obtaining $\mu = 1/3$ and $\beta = 1.3/3\langle k \rangle$, where $\langle k \rangle$ denoted the average degree of the network.

### 3.2   Analysis

To compare the role of different sets of nodes in our simulations, we focus on three measurable features of the epidemic outbreak. First, for $1 \leq d \leq 4$, we measure the empirical spreading power of order $d$ of the node. This is defined as the average number of nodes belonging to the infection subtree rooted in $v$ of depth $d$ (not counting $v$ itself) - i.e., the number of people directly or indirectly infected by $v$. As an example, suppose that the empirical spreading power of order 2 of node $v$ is 30; then, on average over all simulations, 30 people caught the infection either from $v$ or from someone who caught it from $v$.

Second, we gauge the ability of a node to seed a global epidemic, comparing it with its EF value. That is, we estimate the likelihood of observing a global epidemic (i.e. an outbreak in which at least 25% of the population is recovered by the end of the simulation) over multiple independent simulations, as a function of the centrality of the index case.

Finally, we consider the effect of immunizing sets of nodes with a given EF value. That is, we estimate the likelihood of global outbreaks as a function of the average centrality of a set of initially immunized individuals.

### 3.3   Centrality Measures

Centrality measures tackle the fundamental problem of quantifying the *importance* of nodes, based on their topological properties. These measures play a fundamental role in understanding many problems in social network analysis [30], biology [31], urban and power grid design [32, 33], finance [34] and cognitive science [35] among the others.

The geometrical properties captured by a centrality measure can be localized, such as the degree centrality (DC) [36], which ranks the nodes according to how many other nodes they can directly reach. Or they can refer to the global structure of the network, such as betweenness centrality (BC) [37], which counts how many times a node appears in shortest paths between other nodes. Centrality measures, of course, can only capture an instantaneous, static picture of a network, and can account only indirectly for spreading or diffusion processes happening within it [7, 38].

We now briefly define the centrality measures we considered in our analysis - degree centrality, betweenness centrality, and PageRank, alongside Expected Force - highlighting how they relate to spreading processes. While we evaluated many other centrality measures (such as the local clustering coefficient [39]), we only report here on those that best captured the investigated properties of the spreading processes.

**Degree Centrality**  In undirected graphs, the degree centrality (DC) is defined as the number of edges connected to a node, i.e. the node's degree. Nodes with a higher DC can be considered more important in a spreading process, as they can directly infect more nodes.

**Betweenness Centrality**  The betweenness centrality (BC) measures how many times a node appears in a shortest path between other nodes. More precisely, the BC score of a node $v$ in graph $(V, E)$ is defined as

$$BC(v) = \sum_{s,t \in V} \frac{\sigma(s, t|v)}{\sigma(s, t)},$$

where $V$ is the set of nodes in the graph, $\sigma(s, t)$ is the total number of shortest paths between nodes $s$ and $t$, and $\sigma(s, t|u)$ is the number of those shortest paths that pass through node $u$. Nodes with a high betweenness centrality score are considered to be more important or central within the graph, as they have a greater influence on the flow of information or resources through the network.

**Page Rank**  The PageRank ($PG$) metric scores nodes so that nodes connected to important nodes score higher. Specifically, $PG(v)$ of a node $v$ depends on the values in its neighbourhood $N(v)$:

$$PG(v) = \frac{1 - d}{|V|} + \sum_{u \in N(v)} d \frac{PG(u)}{|N(u)|}$$

where $d$ is a parameter (the "damping factor") and $|V|$ is the size of the network. The PageRank metric models the behavior of a random walker traversing the network, with high-scoring nodes more likely to be visited and thus infected during an outbreak [40].

**Expected force**  The Expected Force (EF) tries to capture the *force of infection* that a node exerts on the rest of the network [7]. It does so by looking at all possible sets of nodes infected by that node, on the assumption that the number of their out-going edges will predict their ability to connect the node to the rest of the graph.

The formal definition of Expected Force (EF) depends on a parameter $j$, usually set to 2, and is based on the concept of *transmission cluster*: given a node $v$, a $j$-hops *transmission cluster* $c$ is a tree with exactly $j$ edges rooted at $v$. Intuitively, for an infection that started in $v$, a transmission cluster represents one possible infection state after $j$ transmission events. The *out-degree* $d_c$ of a cluster is the number of edges connecting nodes inside the cluster to nodes outside the cluster - intuitively, it quantifies the ability of the cluster to further propagate the infection. Figure 1 gives an example of two clusters rooted on the same node, and their respective degrees. With $C_v$ we denote the set of all clusters rooted at $v$. For the set $C_v$ of clusters rooted at $v$, and a cluster $c \in C_{v},$, the *normalized cluster degree* $\bar{d}_c$ of $c$ is

$$\bar{d}_c = \frac{d_c}{\sum_{c \in C_v} d_c}.$$

Then, the *expected force* of a node $v \in V$ is the entropy of the distribution of normalized cluster degrees of $C_v$:

$$EF(v) = - \sum_{c \in C_v} \bar{d}_c \log \bar{d}_c. \tag{1}$$

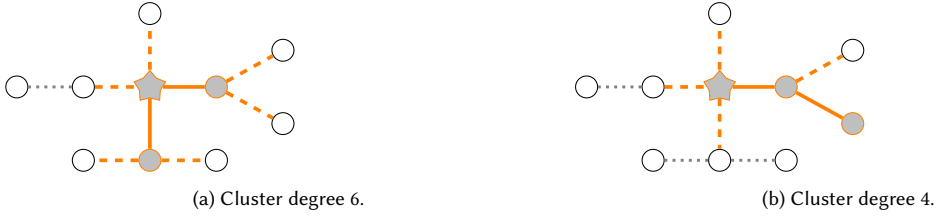

(a) Cluster degree 6.                                          (b) Cluster degree 4.

Fig. 1. Two clusters rooted at the star-shaped node. Gray-shaded nodes belong to the cluster, dashed orange edges connect nodes outside the clusters with nodes inside the clusters. The degree of the cluster corresponds to the number of dashed orange edges.

## 4  PARALLEL COMPUTATION OF EXPECTED FORCE

The original formulation of EF, due to Lawyer [7], calculates Equation (1) for each node in the graph by exploring its neighbourhood and recording the degree of each of its transmission clusters. Computing the EF in this way allows for parallelization at the vertex level, as the computation of the cluster degrees for each vertex is thread-safe and can be performed independently of others. However, this simple approach can lead to workload balancing issues, as the number of clusters each vertex participates in can vary greatly in typical networks. Furthermore, the algorithm is not efficient, as it visits each cluster many times.

To address these drawbacks, we propose an efficient Expected Force algorithm that preserves vertex independence and focuses on the computation of the clusters (*cluster-centric*). The main idea behind the new algorithm is to efficiently enumerate clusters first, then process them independently, recording the cluster degree for each involved node.

Our algorithm, for which pseudocode is provided in Algorithm 1, works in three phases. In the first phase, we consider each node $v$ of the graph in parallel, and collect each pair $i, j$ of its neighbors (line 3). The resulting triplets – the clusters – are accumulated in the set $C$. Since 2-hops clusters always have a node in the middle (not necessarily their seed node), this strategy is ensured to generate all valid transmission clusters. Careful enumeration avoids regenerating the same cluster more than one time. In the second phase, we compute in parallel the degree of each cluster $(i, v, j)$ (line 7) by summing the degree of its nodes and removing 4. If the three nodes form a triangle, the correct cluster degree is obtained by subtracting 2 more edges (line 8). We note here that each cluster is a valid transmission cluster for each of its nodes. Thus, since the cluster degree contributes to the EF of all its components, we can update the cluster size distributions of the three nodes (line 9). In the final phase, we compute for each node the entropy of its cluster size distribution, finally obtaining its expected force of infection (line 12).

Lemma 4.1. *Let $G$ be a graph with $n$ vertices and maximum degree $\delta_{max}$. Algorithm 1 performs $O(n \cdot \delta_{max}^2)$ work and has a critical path of depth $O(\delta_{max}^2)$.*

Proof. In the first phase, the $n$ nodes of the graph are processed in parallel, hence the critical path is determined by the the two nested loops at line 3 that generate the clusters, taking at most $O(\delta_{max}^2)$ time per node. Therefore, the critical path is $O(\delta_{max}^2)$ and the work is $O(n\delta_{max}^2)$.

The second phase works in parallel over the $O(n\delta_{max}^2)$ clusters generated in the first phase, doing a constant number of operations for each cluster, resulting in a $O(1)$ critical path and $(n\delta_{max}^2)$ work. Similarly, the last phase performs a constant number of operations for each node, in parallel, with a $O(1)$ critical path and work $O(n)$.    □

---

**Algorithm 1:** Computing EF$_2$, in parallel

---

**Input:** A graph $G = (V, E)$ of maximum degree $\delta_{max}$
**Output:** The Expected Force

/* $H_v \leftarrow$ Histogram of degrees of the clusters to which $v$ belongs to. With $H_v(d)$ we denote the number of clusters of degree $d$.                                                     */

1  $C \leftarrow$ set of all clusters, initially empty;

   // Phase 1: cluster generation

2  **for** $v \in V$ **do in parallel**

3      **for** $i \in Adj(v)$ **do**

4          **for** $j \in Adj(v), i < j$ **do**

5              $C \leftarrow C \cup \{(i, v, j)\}$;

   // Phase 2: histogram updates

6  **for** $(i, v, j) \in C$ **do in parallel**

7      $d \leftarrow |Adj(v)| + |Adj(i)| + |Adj(j)| - 4$ ;

8      **if** $i \in Adj(j)$ **then** $d \leftarrow d - 2$ ;

9      $H_i(d) \leftarrow H_i(d) + 1$ ;

10     $H_j(d) \leftarrow H_j(d) + 1$ ;

11     $H_v(d) \leftarrow H_v(d) + 2$;

   // Phase 3: computation of the Expected Force

12 **for** $v \in V$ **do in parallel**

13     EF(v) $\leftarrow$ Entropy($H_v$);

---

Note that our parallel algorithm can also be executed sequentially, resulting in a computational complexity of $O(n \cdot \delta_{max}^2)$. This complexity is to be compared with the one of the vertex-centric algorithm[1] originally proposed in [7], which is $O(n \cdot \delta_{max}^2 \cdot \delta_{2,max})$, where $\delta_{2,max}$ is the maximum number of nodes at distance 2 from any node in the graph. The vertex-centric algorithm computes the Expected Force of each node independently from the others: for each node $v$ all its clusters are visited, building the histogram of cluster degrees from which the Expected Force of $v$ is computed. The drawback of this approach is that the same cluster is built and evaluated multiple times, one for each node it contains. This results in the additional $\delta_{2,max}$ factor in the complexity of the algorithm [7].

Our algorithm, instead, visits each cluster just once, updating in the same iteration the histograms of all the nodes contained in the cluster, thus avoiding repeating work and resulting in a better complexity.

## 4.1 Implementation

We implement two algorithms: a parallel version of the vertex-centric approach from Lawyer [7] and our new parallel cluster-centric algorithm, presented in Section 4.

The vertex-centric approach is not suitable for GPU implementation due to the significant workload imbalance, which would result in a large number of idle threads and lead to poor performance. Thus, we implemented it by using OpenMP [41], a multi-platform API for shared-memory parallel programming to exploit traditional multi-core CPUs. We fine-tuned the OpenMP scheduling policy, i.e. the strategy used to assign work to threads in a parallel region. In particular, we adopt dynamic scheduling. With dynamic scheduling, the work is divided into smaller chunks and assigned to threads as they become available. This allows for more flexible and efficient load balancing compared to other scheduling strategies, such as static scheduling, which divides the work into equal parts before the start of the parallel region. Dynamic scheduling also has the advantage of being able to handle irregular and unpredictable workloads, as well as improving performance for loops with small iteration spaces. The chunk size controls the granularity of the work distribution, with smaller chunk sizes leading to more fine-grained load balancing, but also slightly increasing the overhead. In our case, we use chunks of size 2.

---

[1] The original implementation is available at https://github.com/glennlawyer/ExpectedForce.

Differently from the vertex-centric formulation, the cluster-centric algorithm is specifically well-suited for GPU implementation, particularly for the generation of clusters. Our implementation leverages the massive number of threads and the high speed of GPU memory through coalesced memory access by reading the entire edge list stored in Coordinate format (COO) to generate potential clusters. Only for this part of the code, one or more edges are assigned to CUDA threads using an *edge-centric* approach. The algorithm performs a two-step visiting of the graph, resulting in a balanced workload for each vertex. The vertices are then divided into chunks to generate the clusters. After being generated (Phase 1) each cluster is checked to see if it is a triangle, in which case its degree is updated accordingly (Phase 2). This operation is performed per cluster by the GPU by using CUDA streams. CUDA streams are a mechanism for executing multiple tasks on a GPU concurrently and in a specific order. Tasks within the same stream are executed in the order they were enqueued, while tasks in different streams can run concurrently. This approach provides fine-grained control over the cluster to process and maximizes GPU utilization. Moreover, this solution allows to overlap computation (of the GPU) and data movement (between the RAM and GPU memory), as we can process a cluster while the next cluster is being moved from the main memory to the GPU. In the end, each cluster is processed and the EF score of their seeds is updated (Phase 3). Due to space constraints, the discussion on the data structure used to map data and threads on the GPU is omitted. All our implementations are publicly available on a GitHub repository.[2]

## 5 EXPERIMENTAL ANALYSIS

### 5.1 The Networks

We evaluated the effectiveness of EF on a combination of synthetic and real networks. Table 1 summarizes the relevant characteristics of all the graphs used in our study.

The synthetic network for the city of Viterbo is obtained with our Urban Social Network (USN) model [42, 43]. The resulting urban network is unweighted undirected spatial network, in which the probability of an edge connecting two individuals $u$ and $v$ depends on their age, their geographic distance and their sociability. We make use of census data provided by the Italian Institute of Statistics (ISTAT), spatial density estimates provided by the WorldPop Project [44], and aggregated contact data [45] obtained through the SOCRATES Data Tool [46]. Additionally, a vertex–intrinsic social fitness is implemented in the model to control each vertex's sociability, i.e., its tendency to make friends. To obtain a suitable degree distribution, the fitness is taken to be lognormally distributed [42]. USN also consider the real household-structure of the population living in the selected area and individuals are distributed according to the actual population density of the considered territory. The software used to generate these networks is freely available as open-source[3]. This model has been already successfully used to simulate epidemic outbreaks in realistic synthetic urban areas [47, 48]. To assess the efficacy of EF in a wider context, we include in our study real contact/social/interaction networks of varying sizes and features. These are all publicly available[4].

---

| Name | nodes | edges (undirected) | avg. degree | max degree | # of clusters |
|---|---|---|---|---|---|
| **Real Networks** | | | | | |
| contacts | 274 | 4248 | 15 | 101 | 39427 |
| highschool | 327 | 11636 | 36 | 87 | 76999 |
| chees | 7301 | 111798 | 15 | 181 | 862929 |
| social location | 58228 | 428156 | 7 | 1134 | 4474467 |
| Twitter | 465017 | 1667080 | 3 | 667 | 62662902 |
| soc-pocket | 1632803 | 44603928 | 27 | 14854 | 695357852 |
| **Synthetic** | | | | | |
| R-MAT_16,2 | 33753 | 258582 | 2 | 1034 | 2821167 |
| R-MAT_16,4 | 41781 | 513538 | 4 | 1730 | 9971437 |
| R-MAT_16,8 | 49343 | 1018086 | 8 | 2875 | 34342660 |
| R-MAT_16,16 | 55572 | 2012652 | 16 | 4574 | 114937488 |
| R-MAT_16,32 | 60116 | 3969242 | 32 | 6993 | 373831142 |
| Viterbo | 60152 | 603030 | 10 | 78 | 1410085 |

Table 1. Real-world and synthetic networks used in our experiments. An R-MAT with $\approx 2^N$ nodes and $M$ average degree is referred to as R-MAT$_{N,M}$.

We also consider synthetic networks generated according to the R-MAT model [49]. R-MATs networks, which were designed to mimic real-world networks, can be generated efficiently even on large scales, and have been shown to well approximate power-law distributed graphs. In our experiments, we generate R-MATs to study the performance of our implementations in a realistic but controlled setting.

## 5.2 Experimental Setup

The experiments were conducted on a DGX-2 equipped with 8 NVIDIA A100 GPUs, each GPU having 40 GB of memory and 320 Tensor Cores. The system runs on Ubuntu 20.04 LTS and uses CUDA version 11.6. For the experiments on a high-end CPU, we used an AMD EPYC 7742 64-Core Processor and 3 TB of memory. We computed the other centrality measure by using NetworkX [50]. The BC scores of large graphs have been calculated by using an efficient a Multi-GPU implementation of Brandes' algorithm [51].

## 5.3 Performance Analysis

The purpose of our performance analysis is to validate the design and implementation choices described in Section 4 when computing the EF of large-scale graphs. We evaluate our two parallel implementations of EF: the vertex-centric, OpenMP-based implementation of the original EF algorithm [7]; and the cluster-centric, CUDA-based implementation of Algorithm 1. These are the first known parallel implementations for computing the EF and thus deserve a thorough and detailed evaluation. The scalability of the OpenMP-based parallel version will be assessed through both strong scaling (by increasing the number of threads) by measuring the time by varying the average degree of the synthetic graphs. For the CUDA-based implementation of Algorithm 1 we measure the efficiency. To gauge performance, we use the time-to-solution metric (measured in *milliseconds*) and the number of clusters processed per *millisecond*.

Figure 2 shows *strong scaling* results for the OpenMP-based EF algorithm running on an R-MAT graph. The results demonstrate an almost linear relation between parallel workers and running time, indicating no communication bottlenecks. This trend has also been observed in other graphs in our testbed. However, the results depicted in Figure 3 reveal the scalability limitations of the vertex-centric parallel strategy, recording its performance on graphs with growing average degree - as edges are added to a synthetic R-MAT, the OpenMP-based runtime grows much faster than linearly.

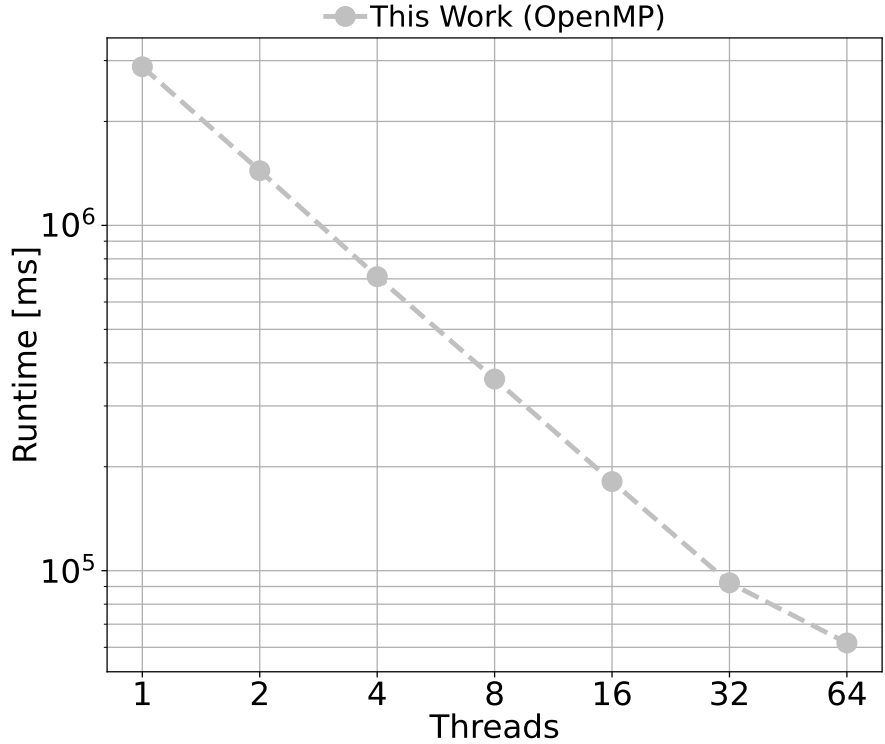

Fig. 2. Strong scaling (time-to-solution) of the vertex-centric algorithm implemented in OpenMP on R-MAT_16,8.

In addition to the scalability issues, the efficiency of the vertex-centric approach is also a concern. The time spent computing each cluster is analyzed in Figure 4. As the number of processed clusters grows, the OpenMP version becomes less efficient, computing multiple time the same cluster, as discussed in Section 4. This behaviour explains its poor performance on larger graphs, since the number of clusters grows much faster than the degree of the graph (refer to the last column in Table 1).

In contrast, the CUDA implementation of the EF algorithm, shown in both Figure 3 and Figure 4, displays a different trend due to its optimized design. Despite the increase in total work (both in terms of edges and clusters), the running time of the CUDA implementation grows at a slower rate than that of the OpenMP, indicating better scaling properties. Furthermore, the cluster-processing rate remains almost constant.

Finally, Figure 5 compares the computation time of all implementations, including the existing sequential version [7], on various graphs listed in Table 1. Due to limited resources, we set a time limit of 24 hours for each computation, and exclude from the plot those computations that exceeded it.

Our findings indicate that the parallel implementations of EF demonstrate substantial advancements compared to the sequential implementation. Among the parallel implementations, the cluster-centric implementation using CUDA stands out as the most efficient, delivering a remarkable maximum speed-up of up to 300× and an average speed-up of 50× compared to the vertex-centric OpenMP implementation. These results showcase the remarkable potential of the CUDA cluster-centric implementation in significantly enhancing the computational efficiency of EF in network analysis.

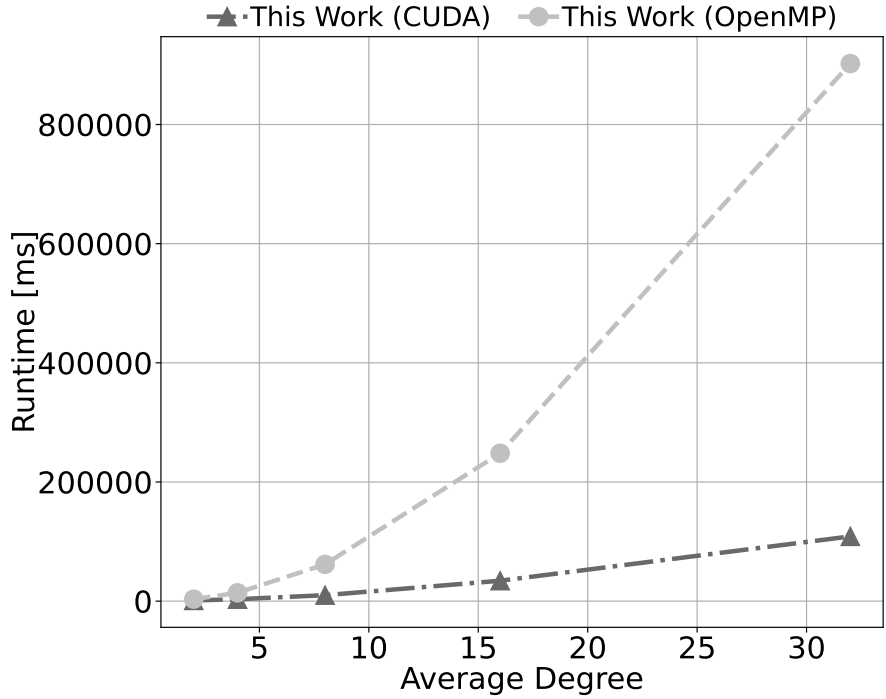

Fig. 3. Comparison (*time to solution*) between the cluster-centric algorithm (Algorithm 1) implemented in CUDA and the vertex-centric OpenMP implementation on a synthetic R-MAT with fixed size ($\approx 2^{16}$) and increasing average degree (2, 4, 8, 16, 32). For this experiment, we fixed the number of threads of the OpenMP implementation to 64.

This high efficiency enabled us to calculate the EF of previously intractable graphs, such as *soc-pocket* - a large-scale contact network counting 1.6 million nodes and 44 million edges. The calculation took less than 4 minutes for the CUDA implementation, compared to almost 18 hours for the OpenMP version, and would have taken days for the sequential algorithm.

### 5.4 Predicting the importance of nodes

In this section, we investigate the empirical relationship between centrality metrics (Section 3.2) and node properties during an epidemic simulation (Section 3.3). In Figure 6, we show Pearson's correlation coefficient between all considered centrality metrics, computed on the static network, and the empirical spreading power, computed based on 100 independent global outbreaks. When considering the Expected Force, we actually take the exponential of its value as, by definition, the EF scales logarithmically with the expected amount of infections "caused" by the considered vertex. With the exception of the graph *social_location*, the EF metric shows equal or better performance than the other centralities in identifying vertices that will be responsible for multiple infection chains.

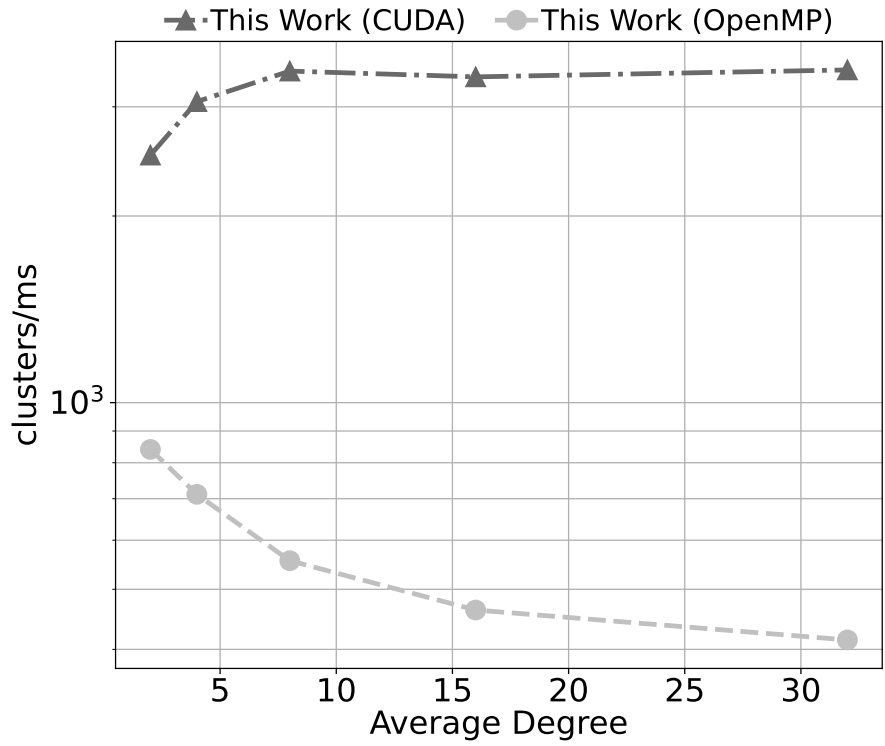

Fig. 4. Comparison (*clusters per millisecond*) between the cluster-centric algorithm (Algorithm 1) implemented in CUDA and the vertex-centric OpenMP implementation on a synthetic R-MAT with fixed size ($\approx 2^{16}$) and increasing average degree ($2, 4, 8, 16.32$). The number of threads used by OpenMP implementation is 64.

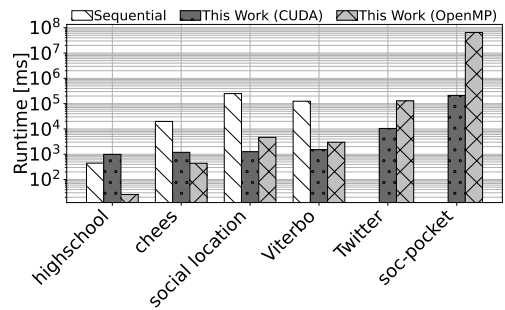

Fig. 5. Running time of three implementations of Expected Force on real-world graphs of increasing size. Graphs details can be found in Table 1.

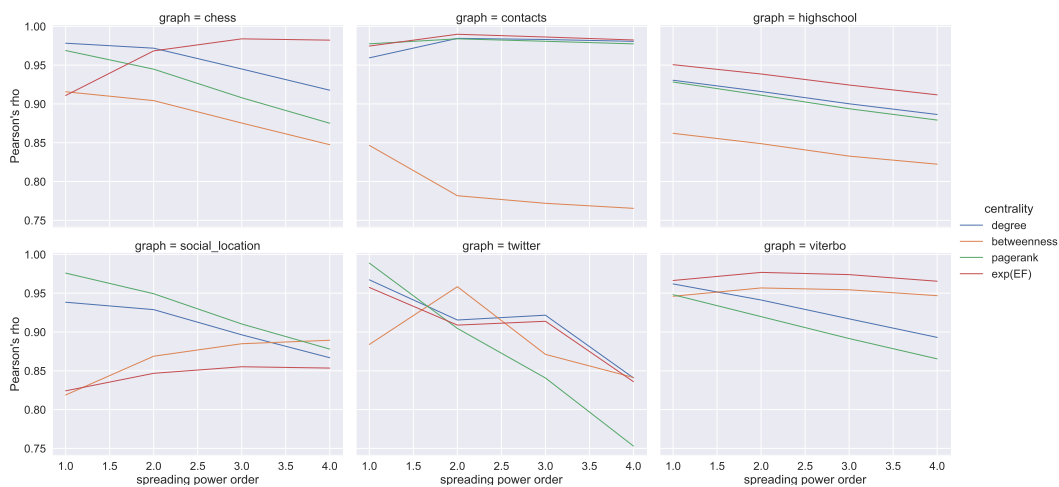

Fig. 6. Pearson's correlation coefficient between the empirical spreading power (of order 1 to 4) and a set of centrality metrics computed on the social graph, for all social graphs considered in this paper.

Next, we study the relationship between EF and global outbreaks. We indentify 10 equally spaced values of the EF and, for each value, we run 100 SIR simulations where a vertex with approximately that centrality value is the index case of the epidemic outbreak. For each such value, in Figure 7 we show the percentage of simulations that gave rise to a global outbreak (> 25% population). Overall, this percentage appears to increase with the EF of the index case. This confirms Lawyer's findings on smaller graphs [7], showing that EF is a good predictor for the ability of a node to start a global outbreak. In the same plot we also show that the final epidemic size increase for increasing values of EF.

Similarly, we simulated epidemic processes where 5% of the network has been immunized, considering 10 different scenarios where the average EF of the immunized vertices varies. In Figure 8 we show the percentage of simulations that resulted in a global outbreak. For most networks, immunizing nodes having large EF produces a significant drop in the likelihood of a major epidemic.

Finally, we assess whether the time course of epidemics is affected by the EF value of the index case. We observe the total length of the epidemic and the time to reach the epidemic peak (*time to peak*). Figure 9 shows that, for all networks except "highschool", both the epidemic length and the time to peak decrease as EF increases. Although preliminary, these results show that individuals with a high EF value have the ability to trigger outbreaks with a higher-than-average probability, spreading the disease faster among their acquaintances, and giving rise to shorter but more severe epidemics.

## 6 CONCLUSION

In this study, we proposed an efficient parallel algorithm for computing the Expected Force, a centrality measure that assesses the relevance of nodes in SIR epidemic models on large networks. Our analysis of the original formulation of Expected Force revealed limitations in terms of scalability and efficiency. To address these limitations, we designed a *cluster-centric* parallel algorithm, with better time complexity and scalability. Our implementation on a NVIDIA GPU showed significant performance improvements with a speed-up of 300× compared to the parallel implementation of the original formulation, allowing us to compute the EF score of a 44 million edges graph in a few minutes. The results of our experiments on networks at different scales demonstrate the applicability of the proposed solution to evaluate the EF in real-world contexts. Comparing the results of extensive SIR simulations with centrality values, we show that EF correlates with spreading power better than other popular, costly centrality metrics such as Betweenness Centrality on a variety of real-world and synthetic networks. Moreover, our results suggest that the EF score is predictive of the ability of nodes to start an epidemic and to hinder it when preemptively immunized.

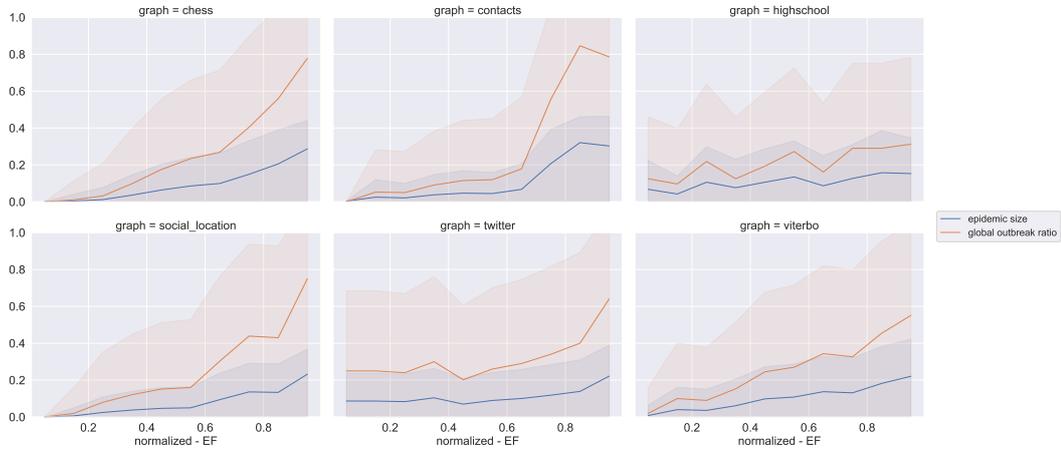

Fig. 7. The proportion of global outbreaks/epidemic size for epidemics starting from vertices having different expected force of infection, for all graphs.

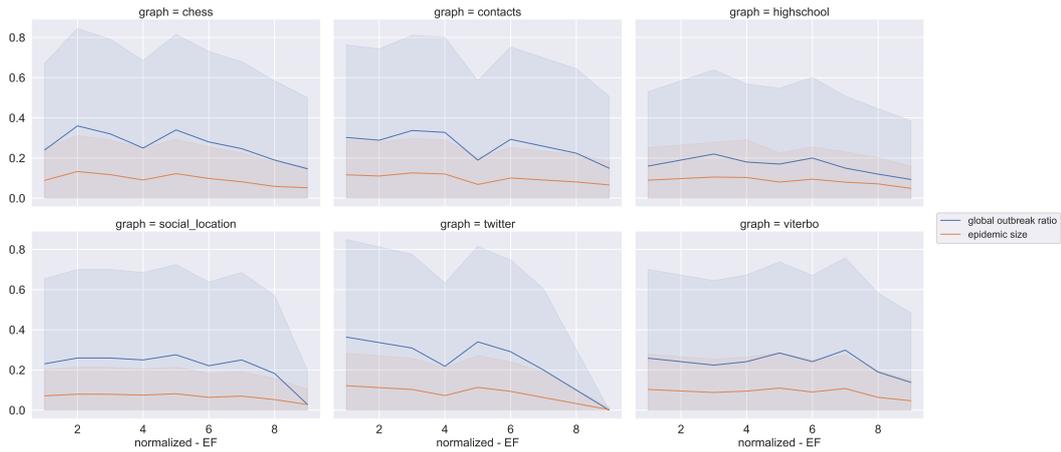

Fig. 8. The proportion of global outbreaks/epidemic size with 5% immunized individuals having different expected force of infection, for all graphs.

The study of centrality measures has gained increased significance with the outbreak of the Covid-19 pandemic. Recent studies have pointed towards the importance of moving towards dynamic or temporal graphs and hypergraphs, as these complex graph structures provide a more comprehensive understanding of network relationships and spreading processes. In this regard, the present study represents a crucial preliminary step towards exploring the Expected Force on these graph structures, serving as a foundation for further research in this area.

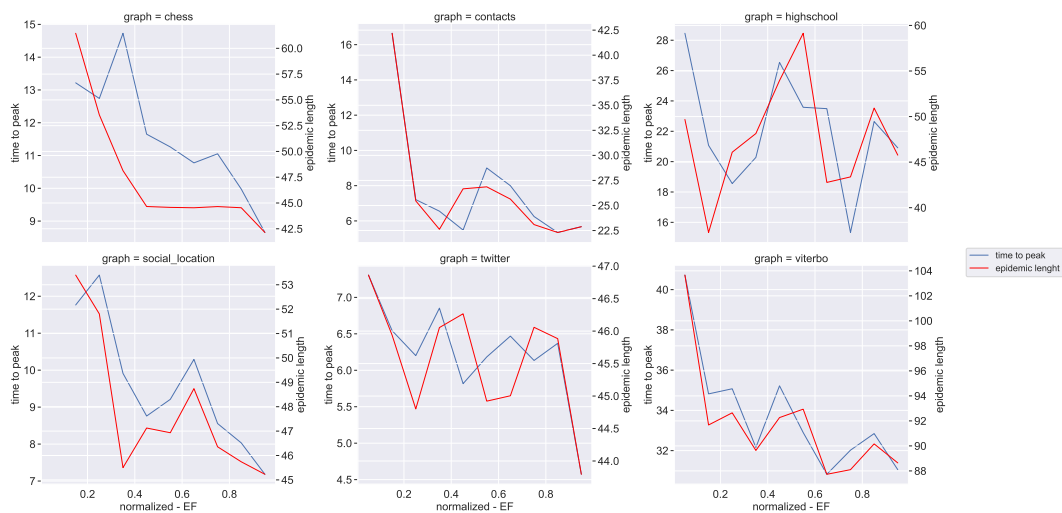

Fig. 9. The mean value of the time to peak/epidemic length versus EF, averaged over 100 global outbreaks, for all graphs.